\documentclass{PoS}
\usepackage{amsmath}
\usepackage{mathrsfs}
\usepackage{bm}

\title{Stability of magnetic condensation and mass generation for confinement in SU(2) Yang-Mills theory}

\ShortTitle{Stability of magnetic condensation and mass generation for confinement in SU(2) Yang-Mills theory}

\author{\speaker{Kei-Ichi Kondo}%
         \thanks{This work is  supported by Grant-in-Aid for Scientific Research (C)  24540252 from Japan Society for the Promotion of Science (JSPS).}\\
        Department of Physics, Faculty of Science, Chiba University, Chiba 263-8522, Japan\\
        E-mail: \email{kondok@faculty.chiba-u.jp}}

\abstract{
In the framework of the functional renormalization group, we reexamine the stability of the Yang-Mills vacuum with a  chromomagnetic condensation. 
We show that the Nielsen-Olesen instability of the Savvidy vacuum with a homogeneous chromomagnetic condensation  disappears in the $SU(2)$ Yang-Mills theory. 
 As a physical mechanism for maintaining the stability even for the small infrared cutoff, we argue that dynamical gluon mass  generation occurs due to a BRST-invariant vacuum condensate of mass dimension-two, which is related to two-gluon bound states identified with glueballs.
These results support the dual superconductor picture for quark confinement. 
}

\FullConference{From quarks and gluons to hadronic matter: A bridge too far?\\
		 2-6 September, 2013\\
		 European Centre for Theoretical Studies in Nuclear Physics and Related Areas (ECT*), Villazzano, Trento (Italy)}

\begin{document}

\section{Introduction}

The \textbf{dual superconductor picture} \cite{dualsuper} for the Yang-Mills theory vacuum is an attractive hypothesis for explaining quark confinement. 
The key ingredients of this picture are the existence of \textbf{chromomagnetic monopole condensation} and the \textbf{dual Meissner effect}.   
For the dual superconductor picture for the Yang-Mills theory vacuum to be true,  the chromomagnetic monopole condensation must  give a more  stable vacuum than the perturbative one. 
In view of this, Savvidy \cite{Savvidy77} has argued based on the general analysis of the renormalization group equation that the  \textbf{dynamical generation of chromomagnetic field} should occur in the Yang-Mills theory, i.e., a non-Abelian gauge theory with asymptotic freedom.  Indeed, Savvidy has shown that the vacuum with non-vanishing  \textbf{homogeneous chromomagnetic field} strength, i.e., the so-called   \textbf{Savvidy vacuum} has lower energy density than the perturbative vacuum with zero  chromomagnetic field.  
The effective potential $V(H)$ of the { homogeneous chromomagnetic field} $H$ has the form  in $SU(2)$ Yang-Mills theory:
\begin{align}
V_{\rm Savvidy}(H)
 =\frac12 H^2
 -\frac{\beta_0 g^2}{16\pi^2}
 \frac12 H^2  \left(\ln\frac{gH}{\mu^2} + c \right)
,\quad
\beta_0 :=-\frac{22}3 <0 .
\end{align}
Then the effective potential $V(H)$ of the homogeneous chromomagnetic field $H$  has an absolute minimum at $H=H_0 \ne 0$ away from $H=0$.
The {chromomagnetic condensation} gives more stable vacuum than the perturbative one. 

Immediately after his proposal, however, N.K.Nielsen and Olesen \cite{NO78} have shown that  the effective potential $V(H)$ of the homogeneous chromomagnetic field $H$, when calculated explicitly at one-loop level in the perturbation theory under the background gauge, develops  a pure imaginary part: 
\begin{equation}
V_{\rm NO}(H)
 =\frac12 H^2
 -\frac{\beta_0 g^2}{16\pi^2} \frac12 H^2
   \left(\ln\frac{gH}{\mu^2} + c  \right)
+   i \ \frac{g^2 H^2}{8\pi}    ,
\end{equation}
in addition to the real part which agrees exactly with  
the Savvidy's result. 
This is called the \textbf{Nielsen-Olesen (NO) instability} of the Savvidy vacuum.
 The presence of the pure imaginary part implies that the Savvidy vacuum gets unstable due to gluon--antigluon pair annihilation. 

This result is easily understood based on the following observation. 
In the homogeneous external chromomagnetic field $H$,  the energy eigenvalue $E_n$ of the massless ({\it off-diagonal}) gluons with the spin $S=1$  ($S_z=\pm 1$)   is given by
\begin{equation}
 E_n^\pm=\sqrt{p_\perp^2+2gH (n+1/2) + 2gH   S_z}  \ (n=0,1,2,\cdots) ,
\end{equation}
where $p_{\perp}$ denotes the momentum in those space-time directions which are not affected by the magnetic field  and the index $n$ is a discrete quantum number which labels the \textbf{Landau levels}.
Then the NO instability is  understood as originating from the \textbf{tachyon mode} with $n=0$ and  $S_z=-1$ (or the lowest Landau level for the gluon with  spin one antiparallel to the external chromomagnetic field), 
since 
\begin{equation}
E_0^-=\sqrt{p_\perp^2-gH} ,
\end{equation}
becomes pure imaginary when $p_\perp^2<gH$.
In other words, the NO instability of the Savvidy vacuum with homogeneous chromomagnetic condensation is due to the existence of the  tachyon mode  corresponding to the lowest Landau level which is realized by the applied external homogeneous chromomagnetic field.

A way to circumvent the NO instability  is to introduce the \textbf{magnetic domains} (domain structure) with a finite extension into the vacuum \cite{Copenhagen}. 
The physical vacuum in Yang-Mills theory is split into an infinite number of domains with macroscopic extensions. 
Inside each such domain there is a nontrivial configuration of  chromomagnetic field and the tachyon mode does not appear in the domain supporting  $p_\perp^2>gH$. 
This resolution for  the NO  instability of Yang-Mills theory  is called the \textbf{Copenhagen vacuum} or \textbf{Spaghetti vacuum}. 
It is instructive to recall that the vortices in a type II superconductor are neatly arranged into a hexagonal or occasionally square lattice.

The Copenhagen vacuum is a well-done model of the Yang-Mills vacuum.  
The domain structure introduces an infrared cutoff which prevents  the momenta from taking the smaller values causing the instability. 
However, it is quite complicated to work out the dynamics of the Yang-Mills theory on the concrete inhomogeneous background. 
Therefore, there have been a lot of works trying to overcome the NO instability for the homogeneous chromomagnetic field.

In view of these, we reexamine the NO instability in the $SU(2)$ Yang-Mills theory  in the framework of the \textbf{functional renormalization group} (FRG)  \cite{Wetterich93} as a realization of the \textbf{Wilsonian renormalization group}.
The FRG enables us to examine the effects caused by changing the infrared cutoff in a systematic way. 
In this paper we follow the methods developed for FRG in \cite{RW94,RW97, LP02,Gies02,EGP11}.
We point out the following results \cite{Kondo13}. 
\begin{enumerate}
\item 
  The \textbf{Nielsen-Olesen instability} in the effective potential $V(H)$ for the homogeneous chromomagnetic field $H$, i.e., the imaginary part Im $V(H)$ of $V(H)$   disappears (or is absent from the beginning) in the framework of the  FRG. 
  (Therefore, the Nielsen-Olesen instability is an artifact   of the one-loop calculation in the perturbation theory and it disappears in the non-perturbative framework beyond the perturbation theory.)

\item 
 However, this result does not necessarily guarantee the automatic existence of the non-trivial homogeneous chromomagnetic field $H_0 \neq 0$ as the minimum of the effective potential $V(H)$, such that 
$
V(H_0) < V(H=0)=0.
$
(Therefore, the absence of the Nielsen-Olesen instability and the existence of the non-trivial minimum for the homogeneous chromomagnetic field in the effective potential are different problems to be considered independently.)

\item 
 As a physical mechanism for maintaining the stability even for the small infrared cutoff, we propose the \textbf{dynamical mass generation} for the off-diagonal gluons (and off-diagonal   ghosts), which is related to the BRST-invariant \textbf{vacuum condensation of mass-dimension two}  \cite{GSZ01,Kondo01,Kondo03,KMSI02}.  
This gives a consistent picture compatible with the absence of the instability. 
(This leads to the \textbf{Abelian dominance} \cite{tHooft81}: in the string tension extracted from the Wilson loop  average  and exponential-falloff of the off-diagonal gluon propagators \cite{AS99,BCGMP03,MCM06} as well as the magnetic monopole dominance  in the Maximal Abelian gauge.)
\end{enumerate}


\section{Complex-valued flow equation in the FRG}


The \textbf{effective average action} $\Gamma_\Lambda$ with the infrared cutoff $\Lambda$ is obtained by solving the flow equation \cite{Wetterich93}: 
\begin{equation}
\partial_{t} \Gamma_{\Lambda} = \frac{1}{2} {\rm STr} \bigl[ ( \Gamma_{\Lambda}^{(2)} + R_{\Lambda} )^{-1} \cdot \partial_{t} R_{\Lambda} \bigr] , \ 
\partial_{t} := \Lambda \frac{d}{d \Lambda} ,
\end{equation}
where ${\rm STr}$ denotes the ``supertrace'' introduced for   writing both  commuting fields (e.g., gluons ) and anticommuting fields (e.g., quarks and the Faddeev-Popov ghosts), $R_\Lambda^\Phi$ is the \textbf{infrared cutoff function} for the field $\Phi$ which is introduced as the \textbf{infrared regulator term}  in the form:
$
\int \Phi^\dagger R_\Lambda^\Phi \Phi ,
$
and $\Gamma_\Lambda^{(2)}$ denotes the second functional derivatives  of $\Gamma_{\Lambda}$ with respect to the field variables $\Phi$: 
\begin{equation}
(\Gamma_\Lambda^{(2)})_{\Phi^\dagger \Phi} = \frac{\overrightarrow{\delta}}{\delta \Phi^\dagger} \Gamma_\Lambda \frac{\overleftarrow{\delta}}{\delta \Phi} ,
\end{equation}
corresponding to the inverse exact propagator at the scale $\Lambda$. 
The ordinary \textbf{effective action} $\Gamma$ as the generating functional of the one-particle irreducible vertex functions is obtained in the limit $\Lambda  \downarrow 0$: $\Gamma = \lim_{\Lambda \downarrow 0} \Gamma_\Lambda$.

We consider the complex-valued effective average action $\Gamma_{\Lambda} = \Gamma_{\Lambda}^{\rm R} + i \Gamma_{\Lambda}^{\rm I}$ which is decomposed into the real part $\Gamma_{\Lambda}^{\rm R}:=\text{Re} \Gamma_\Lambda$ and the imaginary part $\Gamma_{\Lambda}^{\rm I}:=\text{Im} \Gamma_\Lambda$.
Then it is shown (see Appendix A of \cite{Kondo13}) that the flow equation is decomposed into two parts:
\begin{align}
   \partial_{t} \Gamma_{\Lambda}^{\rm R} 
=& \frac{1}{2} {\rm STr} \left\{ \bigl[ ( \Gamma_{\Lambda}^{\rm R (2)} + R_{\Lambda} )^{2} + ( \Gamma_{\Lambda}^{\rm I (2)})^{2} \bigr]^{-1} ( \Gamma_{\Lambda}^{\rm R (2)} + R_{\Lambda} ) \partial_{t} R_{\Lambda}   \right\} , 
\\
   \partial_{t} \Gamma_{\Lambda}^{\rm I} 
=&  - \frac{1}{2} {\rm STr} \left\{ \bigl[ ( \Gamma_{\Lambda}^{\rm R (2)} + R_{\Lambda} )^{2} + ( \Gamma_{\Lambda}^{\rm I (2)})^{2} \bigr]^{-1} \Gamma_{\Lambda}^{\rm I (2)} \partial_{t} R_{\Lambda} \right\} .
\end{align}

We find that \textit{the identically vanishing imaginary part $\Gamma_{\Lambda}^{\rm I}:=\text{Im} \Gamma_\Lambda  \equiv 0 $ is a solution corresponding to a \textbf{fixed point}}:
\begin{equation}
\text{Im} \Gamma_\Lambda \equiv 0  \ \ \text{for any value of} \ \ \Lambda ,
\end{equation}
in sharp contrast with the real part.
If $\Gamma_{\Lambda}^{\rm I} \not= 0$ for a certain value of $\Lambda$, it does not maintain the same value, i.e., $\beta(\Gamma_{\Lambda}^{\rm I})  \not =0$.
Thus the problem of showing the absence of the imaginary part $\text{Im}\Gamma =\lim_{\Lambda \downarrow 0} \text{Im}\Gamma_\Lambda$ in the effective action $ \Gamma=\lim_{\Lambda \downarrow 0} \Gamma_\Lambda$ is reduced to proving the vanishing of the imaginary part  Im$ \Gamma_\Lambda  $ in the effective average action $ \Gamma_\Lambda $ for a sufficiently large value of $\Lambda$:
\begin{equation}
\text{Im} \Gamma_\Lambda  =0 \ \ \text{for a certain value of} \ \ \Lambda \gg 1.
\end{equation}
When $\Gamma_{\Lambda}^{\rm I} = 0$, the flow equation for $\Gamma_{\Lambda}^{\rm R}$ turns into the standard flow equation.


\section{Flow equation in the chromomagnetic background}


We consider the $D$-dimensional Euclidean Yang-Mills theory. 
We decompose the $SU(2)$ Yang-Mills field $\mathscr{A}_\mu=\mathscr{A}_\mu^A T^A$ into the \textbf{background field} $\mathscr{V}_\mu=\mathscr{V}_\mu^A T^A$ and \textbf{the quantum fluctuation field}  $\mathscr{X}_\mu=\mathscr{X}_\mu^A T^A$ where $T^A=\frac{1}{2} \sigma^A$ with $\sigma^A$ being the Pauli matrices ($A=1, 2, 3$):
\begin{equation}
\mathscr{A}_\mu^A =\mathscr{V}_\mu^A + \mathscr{X}_\mu^A  \quad   (A=1, 2, 3)  .
\end{equation}

We can choose without loss of generality    the diagonal field $V_\mu$ as the background field:
\begin{equation}
\mathscr{V}^A_\mu (x) = \delta^{A3} V_\mu (x) ,
\label{V}
\end{equation}
and the off-diagonal field $A_\mu^a$ ($a=1, 2$)  as the quantum fluctuation field:
\begin{equation}
\mathscr{X}_\mu^A (x) =\delta^{Aa} A_\mu^a (x) \ , \ \ (a=1, 2) .
\end{equation}
In what follows, we prepare the diagonal field $V_\mu (x)$ of the form:
\begin{equation}
 V_\mu(x) = \frac{1}{2} x_\nu H_{\nu \mu}  ,
\end{equation}
so that   the  $x$-independent \textbf{homogeneous  background field strength} is realized:
\begin{align}
\mathscr{F}_{\mu \nu}^A [\mathscr{V}](x) 
:=& \partial_\mu \mathscr{V}_\nu^A(x) -\partial_\nu \mathscr{V}_\mu^A(x) +\epsilon^{ABC} \mathscr{V}_\mu^B(x) \mathscr{V}_\nu^C(x)
=  \delta^{A3} \left( \partial_\mu V_\nu(x) -\partial_\nu V_\mu(x) \right) =\delta^{A3} H_{\mu \nu}.
\end{align}

The \textbf{total  effective average action}  $\Gamma_\Lambda$  is specified by giving the gauge-invariant part $\Gamma_\Lambda^{\rm inv}$, the gauge-fixing (GF) part $\Gamma_\Lambda^{\rm GF}$ and the associated Faddeev-Popov (FP) ghost part $\Gamma_\Lambda^{\rm FP}$:
\begin{equation}
\Gamma_\Lambda=\Gamma_\Lambda^{\rm inv} +\Gamma_\Lambda^{\rm GF} + \Gamma_\Lambda^{\rm FP}.
\end{equation}

We choose the \textbf{background gauge}  as the gauge fixing condition to maintain the gauge invariance for the background field. 
In the above choice for the background field (\ref{V}), the background gauge reduces to the \textbf{maximal Abelian} (MA) gauge: 
\begin{equation}
 F^{a} := \mathscr{D}_\mu^{ab} [V] A_\mu^b =0 , 
\quad
\mathscr{D}_\mu^{ab}[V]  := \partial_\mu \delta^{ab}-g \epsilon^{ab3} V_\mu .
\end{equation}
Then  \textbf{the gauge-fixing term} is given by
\begin{equation}
\Gamma^{\rm GF} =\int d^Dx \frac{1}{2\alpha} \left( \mathscr{D}_{\mu}^{ab} [V] A_\mu^b \right)^2 , 
\end{equation}
where $\alpha$ denotes the gauge-fixing parameter. 

The \textbf{FP ghost term} is determined according to the standard procedure (see e.g., \cite{Kondo97}) as 
\begin{align}
\Gamma^{\rm FP} =&  \int d^Dx  \{ 
  i \bar{C}^{a} \mathscr{D}_{\mu}^{ab} [V] \mathscr{D}_{\mu}^{bc} [V] C^{c} 
- g^{2} \epsilon^{ab3} \epsilon^{cd3}  i\bar{C}^{a} C^{d} A_{\mu}^{b} A_{\mu}^{c}  
+ i \bar{C}^{a} g\epsilon^{ab3} (\mathscr{D}_{\mu}^{bc} [V] A_{\mu}^{c}) C^{3} \} .
\end{align}

For the gauge-invariant part $\Gamma_\Lambda^{\rm inv}$, we adopt the  ansatz,  a function $W_\Lambda$  of the gauge-invariant term $\Theta$ constructed from the field strength $ \mathscr{F}_{\mu \nu}^A [\mathscr{A}]:= \partial_\mu \mathscr{A}_\nu^A-\partial_\nu \mathscr{A}_\mu^A+\epsilon^{ABC} \mathscr{A}_\mu^B \mathscr{A}_\nu^C $:
 \begin{equation}
\Gamma_\Lambda^{\rm inv} =\int d^Dx W_\Lambda \left( \Theta (x) \right) , 
\quad
\Theta:= \frac{1}{4} \left( \mathscr{F}_{\mu \nu}^A [\mathscr{A}] \right)^2 .
\end{equation}
 $\Theta$ is decomposed as  \cite{Kondo10}
\begin{align}
\Theta =& \frac{1}{4} \left( \mathscr{F}_{\mu \nu}^A [\mathscr{V}] \right)^2 +
 \frac{1}{2} A^{\mu a} \left( Q_{\mu \nu}^{ab}+\mathscr{D}_\mu^{ac}[V] \mathscr{D}_\nu^{cb} [V] \right) A^{\nu b} 
+\frac{1}{4} \left( \epsilon^{3ab} A_\mu^a A_\nu^b \right)^2 ,
\end{align}
\begin{align}
  Q_{\mu \nu}^{ab} :=& - \left(\mathscr{D}^2 \right)^{ab} \delta_{\mu \nu} +2g \epsilon^{ab} H_{\mu \nu} ,
\quad
\left(\mathscr{D}^2 \right)^{ab} :=  \mathscr{D}_\rho^{ac} [V] \mathscr{D}_\rho^{cb} [V] .
\end{align}
In the vanishing off-diagonal field limit $A_\mu^a \to0$, $\Theta$ is reduced  to
\begin{align}
\Theta|_{A=0}=& \frac{1}{4} \left( \mathscr{F}_{\mu \nu}^A [\mathscr{V}](x) \right)^2 
=\frac{1}{4} \left( \partial_\mu V_\nu(x)  - \partial_\nu V_\mu(x)  \right)^2
=  \frac{1}{2} H^2 ,
\end{align}
where 
$
 H :=   \sqrt{\bm{H}^2} = \sqrt{\frac{1}{2} H_{\alpha \beta} H_{\alpha \beta}}>0  .
$

The off-diagonal gluon fields $A_\mu^a$ (and  off-diagonal ghost fields $C^a$, $\Bar{C}^a$) should be integrated out in the framework of the FRG following the idea of  the Wilsonian renormalization group.
For this purpose, we introduce the \textbf{infrared regulator term} $\Delta S_\Lambda$ for the off-diagonal gluon $A_\mu^a$ and off-diagonal ghosts $C^a$, $\Bar{C}^a$ by
\begin{align}
\Delta S_\Lambda =& \int_p \Big[ \frac{1}{2} A_\mu^a (p) R_{\Lambda, \mu \nu} (p^2) \delta^{ab} A_\nu^b (p) 
+\Bar{C}^a (p) R_\Lambda (p^2) \delta^{ab} C^b (-p) \Big] \ (a,b=1,2),
\end{align}
where 
$
 \int_p := \int \frac{d^Dp}{(2\pi)^D}
$
  denotes the integration over the $D$-dimensional momentum space.
We choose  the infrared cutoff function with the structure:
\begin{equation}
R_{\Lambda, \mu \nu} (p^2) = \delta_{\mu \nu}  R_\Lambda (p^2) .
\end{equation}

We adopt the \textbf{proper-time form} of the flow equation \cite{LP02}:
\begin{equation}
\partial_t \Gamma_\Lambda =\int^\infty_0 d\tau \frac{1}{2} {\rm STr} \left[ e^{-\tau \left(\Gamma_\Lambda^{(2)}+R_\Lambda \right)} \partial_t R_\Lambda \right].
\end{equation}
After performing the mode decomposition according to the projection method \cite{RW94,RW97, Gies02,EGP11},   the flow equation reads
\begin{align}
\partial_{t} \Gamma_{\Lambda} =& \frac{1}{2} \int_{0}^{\infty} d \tau \ \Omega^{-1} {\rm Tr} \bigl[ e^{- \tau ( W_{\Lambda}^{\prime} Q + R_{\Lambda}^{\rm gluon} )} \cdot \partial_{t} R_{\Lambda}^{\rm gluon} \bigr]  (\text{transverse gluons})
\nonumber\\&
- \frac{1}{2} \int_{0}^{\infty} d \tau \ \Omega^{-1} {\rm Tr} \bigl[ e^{- \tau (  - W_{\Lambda}^{\prime} \mathscr{D}^2 + R_{\Lambda}^{\rm gluon} )} \cdot \partial_{t} R_{\Lambda}^{\rm gluon} \bigr]  (\text{scalar gluon})
\nonumber\\&
+ \frac{1}{2} \int_{0}^{\infty} d \tau \ \Omega^{-1} {\rm Tr} \bigl[ e^{- \tau (  -\alpha_\Lambda^{-1}\mathscr{D}^2 + R_{\Lambda}^{\rm gluon} )} \cdot \partial_{t} R_{\Lambda}^{\rm gluon} \bigr]   (\text{longitudinal gluon})
\nonumber\\&
- \int_{0}^{\infty} d \tau \ \Omega^{-1} {\rm Tr} \bigl[ e^{- \tau ( - \tilde{Z}_{\Lambda} \mathscr{D}^2 + R_{\Lambda}^{\rm ghost} )} \cdot \partial_{t} R_{\Lambda}^{\rm ghost} \bigr]   (\text{ghosts })
 ,
\end{align}
The spectrum sum is obtained from eigenvalues of the respective operator. 
The \textbf{covariant Laplacian} $- \left( \mathscr{D}_{\rho} [\mathscr{V}] \right)^{2}$ with the background field $\mathscr{V}$ which gives the (covariant constant) uniform chromomagnetic field $H$ has the spectrum:
\begin{equation}
{\rm Spect} \bigl[ - \mathscr{D}_{\rho}^{2} [\mathscr{V}] \bigr] = p_{\perp}^{2} + ( 2 n + 1 ) gH , \ 
( n = 0 , 1 , \cdots ),
\end{equation}
where $p_{\perp}$ denotes the $(D-2)$ dimensional (Fourier) momentum in those space-time directions which are not affected by the magnetic field (say, orthogonal to $1-2$ plane) and the index $n$ is a discrete quantum number which labels the Landau levels.
We take into account the fact that the density of states is $\frac{gH}{2\pi}$ for the \textbf{Landau levels}.

Moreover, the operator $Q_{\mu\nu}^{ab}$ with the same background field $\mathscr{V}$ has the spectrum:
\begin{align}
{\rm Spect} \bigl[ Q_{\mu\nu}^{ab} \bigr] =&
\left\{ \begin{array}{c}
p_{\perp}^{2} + ( 2 n + 1 ) gH \\
p_{\perp}^{2} + ( 2 n + 3 ) gH \\
p_{\perp}^{2} + ( 2 n - 1 ) gH
\end{array} \right.   {\rm multiplicity} \ \left. \begin{array}{c}
(D - 2) \\
1 \\
1 
\end{array} \right. 
( n = 0 , 1 , \cdots ),
\end{align}
where the last term contains the \textbf{Nielsen-Olesen unstable mode}  for $n=0$, i.e., 
\begin{equation}
 p_{\perp}^{2}-gH , 
\end{equation}
which becomes a \textbf{tachyonic mode} for small momenta $p_{\perp}^{2} < gH$.

where $W'_\Lambda (\Theta) = \frac{d}{d\Theta} W_\Lambda (\Theta)$.
Here we have introduced  the \textbf{wavefunction renormalization constants}:
$
Z_\Lambda =Z_\Lambda^{\rm gluon} \ , \ \Tilde{Z}_\Lambda=Z_\Lambda^{\rm ghost} .
$
In this derivation, we have adopted the truncation: neglecting the four-point interactions 
among the off-diagonal gluons and off-diagonal ghosts 
$
- g^{2} \epsilon^{ab3} \epsilon^{cd3}  i\bar{C}^{a} C^{d} A_{\mu}^{b} A_{\mu}^{c}  
$, 
which do not couple to the background field $V_\mu$.

The respective trace without the infrared regulator $R_\Lambda$ is easily obtained:
\begin{align}
  \Omega^{-1} {\rm Tr} [ e^{- \tau ( W_{\Lambda}^{\prime} Q ) } ]
=& \frac{NgH}{(4\pi)^{\frac{D}{2}}} ( \tau W_{\Lambda}^{\prime} )^{1 - \frac{D}{2}}    \frac{ 2(D-2)  e^{-\tau W_{\Lambda}^{\prime} gH} + 2 e^{-3\tau W_{\Lambda}^{\prime} gH} +  2 { e^{\tau W_{\Lambda}^{\prime} gH} }  }{1-e^{-2\tau W_{\Lambda}^{\prime} gH}}  
 ,
\nonumber\\
\Omega^{-1} {\rm Tr} [ e^{- \tau ( -W_{\Lambda}^{\prime} \mathscr{D}^2 ) } ]
=& \frac{NgH}{(4\pi)^{\frac{D}{2}}} ( \tau W_{\Lambda}^{\prime} )^{1 - \frac{D}{2}}  \biggl[ \frac{2 e^{-\tau W_{\Lambda}^{\prime} gH} }{ 1-e^{-2\tau W_{\Lambda}^{\prime} gH} }  \biggr] ,
\nonumber\\
\Omega^{-1} {\rm Tr} [ e^{- \tau ( -\tilde{Z}_{\Lambda}  \mathscr{D}^2 ) } ]
=& \frac{NgH}{(4\pi)^{\frac{D}{2}}} ( \tau \tilde{Z}_{\Lambda} )^{1 - \frac{D}{2}}  \biggl[ \frac{ 2 e^{-\tau \tilde{Z}_{\Lambda} gH} }{1-e^{-2\tau \tilde{Z}_{\Lambda} gH} }  \biggr] ,
\end{align}
where $N=2$ for $SU(2)$.

In order to  obtain the closed analytical form for the solution and to compare the FRG calculations with the loop calculations, 
we choose the momentum-independent \textbf{infrared regular of the mass type}:
\begin{equation}
R_\Lambda^\Phi=Z_\Lambda^\Phi \Lambda^2 ,
\label{mass-reg}
\end{equation}
where $Z_\Lambda^\Phi$ denotes the \textbf{wave function normalization}  constant for the field $\Phi$. 
Moreover, we show later that the result is independent of the choice of the infrared regulator.

For the infrared regulator of the mass type, thus the flow equation reads
\begin{align}
\partial_{t} \Gamma_{\Lambda}  
 =&  \frac{NgH}{(4\pi)^{\frac{D}{2}}}  \Big\{  ( W_{\Lambda}^{\prime} )^{1- \frac{D}{2}}  ( 2 - \eta_{\Lambda} ) Z_{\Lambda} \Lambda^{2}  
\int_{0}^{\infty} d \tau \tau^{1 - \frac{D}{2}} e^{- \tau Z_{\Lambda} \Lambda^{2}}   
\frac{ (D-2)e^{-\tau W_{\Lambda}^{\prime}gH}+e^{-3\tau W_{\Lambda}^{\prime}gH}+e^{\tau W_{\Lambda}^{\prime}gH}}{1-e^{-2\tau W_{\Lambda}^{\prime}gH}}   
\nonumber\\&
-   ( W_{\Lambda}^{\prime} )^{1- \frac{D}{2} }  ( 2 - \eta_{\Lambda} ) Z_{\Lambda}  \Lambda^{2}  
\int_{0}^{\infty} d \tau \tau^{1 - \frac{D}{2}}  e^{- \tau Z_{\Lambda} \Lambda^{2}}  \frac{  e^{-\tau W_{\Lambda}^{\prime} gH} }{1-e^{-2\tau W_{\Lambda}^{\prime} gH}} 
\nonumber\\&
+   \alpha_\Lambda^{ \frac{D}{2}-1}  ( 2 - \eta_{\Lambda} ) Z_{\Lambda}  \Lambda^{2}  
\int_{0}^{\infty} d \tau \tau^{1 - \frac{D}{2}}  e^{- \tau Z_{\Lambda} \Lambda^{2}}  \frac{  e^{-\tau \alpha_\Lambda^{-1} gH} }{1-e^{-2\tau \alpha_\Lambda^{-1} gH}} 
\nonumber\\&
 - (\tilde{Z}_{\Lambda})^{1-\frac{D}{2}} (2 - \tilde{\eta}_\Lambda) \tilde{Z}_{\Lambda} \Lambda^{2} 
\int_{0}^{\infty} d \tau \tau^{1 - \frac{D}{2}}  e^{- \tau \tilde{Z}_{\Lambda} \Lambda^{2}}  \frac{2e^{-\tau \tilde{Z}_{\Lambda} gH}}{1-e^{-2\tau \tilde{Z}_{\Lambda} gH}}  \Big\} ,
\label{flow-eq1}
\end{align}
where we have introduced  the \textbf{anomalous dimensions}:
\begin{align}
\eta_\Lambda &:=-\partial_t \ln Z_\Lambda =-Z_\Lambda^{-1} \partial_t Z_\Lambda , 
\quad
\Tilde{\eta}_\Lambda  :=-\partial_t \ln \Tilde{Z}_\Lambda =-\Tilde{Z}_\Lambda^{-1} \partial_t \Tilde{Z}_\Lambda.
\end{align}

We find that the integral with respect to $\tau$ on the right-hand side of the flow equation  is divergent at $D=4$ in the $\tau=0$ region which is an \textbf{ultraviolet divergence}. 
This divergence is independent of the \textbf{infrared  divergence} coming from $\tau=\infty$ region due to the factor $e^{\tau W_{\Lambda}^{\prime}gH}$ for which the Nielsen-Olesen instability is responsible.
This ultraviolet divergence is due to the fact that  the momentum-independent infrared cutoff function of the mass type  does not suppress the high-momenta.
This aspect is a short-coming of the mass-type infrared regulator. 
However, the result will be true for any other choice of the infrared regulator, since the infrared regulator $R_\Lambda (p^2)$ is constructed so that  any infrared cutoff function approaches the same asymptotic form as the mass-type one in the large $\Lambda$.
See \cite{Kondo13} for more details.

The ultraviolet divergence is removed by the standard method.
Thus we arrive at the flow equation without the ultraviolet divergence:
\begin{align}
\partial_{t} \Gamma_{\Lambda}   
=&  \frac{N}{2} \frac{2gH }{(4\pi)^{2 }} 
  \left(  -\ln \frac{2gH}{4 \pi \mu^2 } - \gamma \right) 
\Biggr\{ ( W_{\Lambda}^{\prime} )^{-1}  ( 2 - \eta_{\Lambda} ) Z_{\Lambda} \Lambda^{2}  
\nonumber\\&  
\times  
    \Bigg[   \zeta \left( 0, \frac{1}{2} +\frac{Z_{\Lambda} \Lambda^{2}}{2 W_{\Lambda}^{\prime}gH}  \right) 
    +\zeta \left( 0, \frac{3}{2} +\frac{Z_{\Lambda} \Lambda^{2}}{2 W_{\Lambda}^{\prime}gH}  \right) 
 + \zeta \left( 0, -\frac{1}{2} +\frac{Z_{\Lambda} \Lambda^{2}}{2 W_{\Lambda}^{\prime}gH}  \right)   \Bigg] 
\nonumber\\ &
   + \alpha_\Lambda   ( 2 - \eta_{\Lambda} ) Z_{\Lambda}  \Lambda^{2}     \zeta \left( 0, \frac{1}{2} +\frac{Z_{\Lambda} \Lambda^{2}}{2 \alpha_\Lambda^{-1} gH}  \right)
 - 2 (2 - \tilde{\eta}_\Lambda) \Lambda^{2}  \zeta \left( 0, \frac{1}{2} +\frac{ \Lambda^{2}}{2 gH}  \right)   \Biggr\}
\nonumber\\ & 
 +  \frac{N}{2} \frac{2gH }{(4\pi)^{2 }} 
 \Biggr\{  (W_{\Lambda}^{\prime} )^{-1}  ( 2 - \eta_{\Lambda} ) Z_{\Lambda} \Lambda^{2}  
 \Bigg[   \zeta^{(1,0)} \left( 0, \frac{1}{2} +\frac{Z_{\Lambda} \Lambda^{2}}{2 W_{\Lambda}^{\prime}gH}  \right) 
 +\zeta^{(1,0)} \left( 0, \frac{3}{2} +\frac{Z_{\Lambda} \Lambda^{2}}{2 W_{\Lambda}^{\prime}gH}  \right)      
\nonumber\\ &  \quad\quad
  + \zeta^{(1,0)} \left( 0, -\frac{1}{2} +\frac{Z_{\Lambda} \Lambda^{2}}{2 W_{\Lambda}^{\prime}gH}  \right)   
-2 \zeta \left( 0, \frac{1}{2} +\frac{Z_{\Lambda} \Lambda^{2}}{2 W_{\Lambda}^{\prime}gH}  \right)
 \Bigg]   
\nonumber\\&
 \quad\quad +   \alpha_\Lambda   ( 2 - \eta_{\Lambda} ) Z_{\Lambda}  \Lambda^{2}    \zeta^{(1,0)}\left( 0, \frac{1}{2} +\frac{Z_{\Lambda} \Lambda^{2}}{2 \alpha_\Lambda^{-1} gH}  \right)  
- 2   (2 - \tilde{\eta}_\Lambda)   \Lambda^{2}   \zeta^{(1,0)}\left( 0, \frac{1}{2} +\frac{ \Lambda^{2}}{2 gH}  \right)     \Biggr\} .
\label{flow-eq5}
\end{align}
Here we have introduced the \textbf{generalized Riemann $\zeta$-function} or the \textbf{Hurwitz $\zeta$-function} $\zeta (z, \lambda)$ defined by 
\begin{equation}
\zeta (z, \lambda) :=\sum \limits_{n=0}^{\infty} \frac{1}{(n+\lambda)^z} ,
\end{equation}
which has its integral representation
\begin{equation}
\zeta (z, \lambda) = \frac{1}{\Gamma(z)} \int^\infty_0 ds \ s^{z-1} \frac{e^{-\lambda s}}{1-e^{-s}} \ \ ( \text{Re} z>1 , \text{Re} \lambda>0 ) .
\end{equation}
Although the Hurwitz $\zeta$-function $\zeta (z , \lambda)$ is originally defined for ${\rm Re} \ z > 1 , \ {\rm Re} \ \lambda > 0$, 
it can be analytically continued to other region in the complex $z$-plane as an analytic function.



\section{Absence of the Nielsen-Olesen instability}


For large $\Lambda$, we can take the approximation:
\begin{align}
&\text{i}) \   W_\Lambda (\Theta)  = \Theta \Rightarrow W^\prime_\Lambda (\Theta) \equiv 1 
   \Longleftrightarrow Z_\Lambda \equiv 1 \Rightarrow \eta_\Lambda \equiv 0 , 
 \nonumber\\
&\text{ii)} \  \Tilde{Z}_\Lambda \equiv 1 \Rightarrow  \Tilde{\eta}_\Lambda \equiv 0 ,
 \nonumber\\
&\text{iii)} \  \alpha_\Lambda \equiv\alpha_{\Lambda_{\rm UV}} = \text{const.} \ge 0 .
\end{align}
Then the flow equation can be cast into the total derivative form:
$
 \partial_{t} \Gamma_{\Lambda}   
=  \partial_{t} (...) .
$
We take into account the fact that the effective average action  $\Gamma_{\Lambda}$ at $\Lambda=\Lambda_{\rm UV}=\infty$ is given by the bare action for the classical chromomagnetic  field   background:
$
 \Gamma_{\Lambda=\infty} =\frac{1}{4} \left( \mathscr{F}_{\mu \nu}^A [\mathscr{V}] \right)^2 =\frac{1}{2} H^2 .
$
Then an approximate solution is obtained by integrating the flow equation from $\Lambda=\Lambda_{\rm UV}=\infty$ to $\Lambda$ 
where $\Tilde{V}_\Lambda (H)=0$ at $\Lambda=\infty$.

Thus, we arrive at the effective potential for large $\Lambda$, e.g., in the case of $\alpha_\Lambda \equiv 1$:
\begin{align}
V_\Lambda (H)  &=  \frac{1}{2} H^2 +\frac{1}{16\pi^2} \Lambda^2 \left[ \ln \frac{gH}{\mu^2} +\frac{1}{4}-C \right] 
- \frac{2}{16\pi^2} gH \Lambda^2 \ln \frac{\Lambda^2-gH}{\Lambda^2+gH} \notag \\
&+ \frac{1}{16\pi^2} g^2 H^2 \Bigg[ \frac{11}{3} \ln \frac{gH}{\mu^2} 
+ 2 \ln  \frac{\Lambda^2 +gH}{gH}  +  2 \ln \frac{\Lambda^2 -gH}{gH}   
 \notag \\ &
  -\frac{11}{3}C -4 \ln 2 -\frac{1}{3} +8 \zeta^{(1,0)} \left( -1, \frac{1}{2}+ \frac{\Lambda^2}{2gH} \right) \Bigg ] .
\label{solution}
\end{align}
Note that $\zeta^{(1,0)}(-1,\lambda)$ is real-valued for $\lambda>0$. 

For the large $\Lambda$ satisfying  $\Lambda^2 \ge gH$, $\Tilde{V}_\Lambda (H)$ is real-valued and $V_\Lambda (H)$ has no imaginary part:
\begin{equation}
\text{Im} V_\Lambda (H) =0  \ \text{for} \ \Lambda^2 \ge gH ,
\end{equation}
\begin{equation}
  \partial_t \text{Im} V_\Lambda (H) = 0 \ \text{for} \ \Lambda^2 \ge gH  .
\end{equation}
Therefore, the Nielsen-Olesen instability disappears for any value of $\Lambda$, in particular even at $\Lambda=0$ according to the above argument  of the fixed point for the pure imaginary part of the flow equation. 

For the small $\Lambda$ satisfying $\Lambda^2 <gH$, however, the effective average potential $V_\Lambda (H)$ obtained above has the non-vanishing imaginary part:
\begin{align}
  \text{Im} V_\Lambda (H) &= \frac{4}{16\pi^2} g^2 H^2 \frac{\frac{\Lambda^2}{gH}-1}{2} \ln (-1)/i 
=\frac{1}{8\pi} gH (gH-\Lambda^2)    \ \ \text{for} \ \Lambda^2 <gH ,
\end{align}
which yields the nontrivial flow of the imaginary part:
\begin{equation}
\partial_t \text{Im} V_\Lambda (H) = - \frac{1}{4\pi} gH \Lambda^2 < 0 \ \ \text{for} \ \Lambda^2<gH .
\end{equation}
This is not a contradiction, since the approximate solution of $V_\Lambda (H)$ obtained above is not considered to be valid in the small $\Lambda$ region; $\Lambda^2 <gH$. 
In fact, the derivative $\partial_t \text{Im} V_\Lambda (H)$ has the discontinuity at $\Lambda^2=gH$.
The effective potentials obtained above  reproduce the Nielsen-Olesen result by putting $\Lambda = 0$.

 
\section{gluon mass generation and vacuum condensations}


 The above approximate solution (\ref{solution}) eventually has the imaginary part and hence cannot be used in the limit $\Lambda \to 0$. 
As will be shown in this section, however, the approximate solution obtained in the same type of approximations has the limit $\Lambda \to 0$ without developing the imaginary part, if the effects of mass generation are incorporated into the analysis. 
Such mass generation is expected to occur, as established in the numerical simulations on the lattice \cite{AS99,BCGMP03}. 

We introduce  the mixed composite operators of gluons and ghosts: 
For $SU(2)$,  
\begin{equation}
\mathcal{O}=\frac{1}{2} A_\mu^a A^{\mu a} +\alpha i \Bar{C}^a C^a \ \ (a=1, 2).
\end{equation}
and study the mass generation for the off-diagonal gluons (and ghosts), originating from the dimension-two condensation $\langle \mathcal{O} \rangle$.
It is shown \cite{Kondo01} that the dimension-two condensation $\langle \mathcal{O} \rangle$ is BRST invariant %
 in the modified MA gauge  \cite{Kondo98} defined by  the GF+FP term, 
i.e., the $OSp(D,2)$-invariant form:
\begin{align}
\mathscr{L}_{\rm GF+FP}^{\rm MA} 
=&  i \bm{\delta} \bar{\bm{\delta}} \left( \frac{1}{2}  A_{\mu}^{a} A_{\mu}^{ a} + \frac{\alpha}{2} i \bar{C}^{a} C^{a} \right) 
 ,
\end{align}
where $\bm{\delta}$ and $\bar{\bm{\delta}}$ are respectively the BRST and anti-BRST transformations.

According to \cite{EW94}, we introduce a new field $\phi$ which is an auxiliary field  with no kinetic term represented by the Lagrangian density:
\begin{align}
\mathscr{L}_\phi
 =\frac{1}{2} \left( \phi+G\mathcal{O} \right)^\dagger G^{-1} \left( \phi +G\mathcal{O} \right) 
 =\frac{1}{2} \phi^\dagger G^{-1} \phi +\phi^\dagger \mathcal{O} +\frac{1}{2} \mathcal{O}^\dagger G \mathcal{O} ,
\end{align}
by inserting the unity: 
$
1=\int D\phi e^{-\int d^Dx \mathscr{L}_\phi} ,
$
in the path-integral measure.
We observe: 
\begin{itemize}
\item
 From the first term $\frac{1}{2} \phi^\dagger G^{-1} \phi$, we observe that $G$ represents the effective propagator of the \textbf{collective field} $\phi$, i.e., \textbf{ two-gluon  bound state} propagator. 

\item
The second term $\phi^\dagger \mathcal{O}$ yields the cubic interactions $\phi A A$ (and $\phi \Bar{C} C$) for the operator $\mathcal{O}$ quadratic in the off-diagonal gluons (and ghosts). 

\item
The third term $\frac{1}{2} \mathcal{O}^\dagger G \mathcal{O}$ involving only the fundamental fields has the form of an exchange of $\phi$ in the tree approximation. 
\end{itemize}

By including $\mathscr{L}_\phi$, the two-point functions $\Gamma_\Lambda^{(2)}$ are modified as
\begin{align}
\left( \Gamma_\Lambda^{(2)} \right)_{A_\mu^a A_\nu^b} &= W_\Lambda^\prime Q_{\mu \nu}^{ab} +\varphi \delta_{\mu \nu} \delta^{ab} , 
\quad
\left(\Gamma_\Lambda^{(2)} \right)_{\Bar{C}^a C^b}   =  -\Tilde{Z}_\Lambda \left( \mathscr{D}^2 \right)^{ab}+\alpha_\Lambda \varphi \delta^{ab} ,
\end{align}
where 
$
  \varphi=\langle \phi \rangle .
$
Here we have adopted the truncation: neglecting the four-point interactions 
among the off-diagonal gluons and off-diagonal ghosts.

We use the infrared regulator of the mass type and the same approximations for $W_\Lambda$, $\Tilde{Z}_\Lambda$ and $\alpha_\Lambda$ as those adopted in the previous case. Then we obtain the effective average potential $V_\Lambda (H, \varphi)$ describing the chromomagnetic   condensation and dynamical mass generation simultaneously.
We consider the simplest case of $\alpha_\Lambda \equiv 1$ to clarify  the qualitative feature   (see \cite{Kondo03} for a physical meaning of the dimension-two condensate in this gauge).
In this case, the effective potential is given by
\begin{align}
V_\Lambda (H, \varphi)  =& \frac{1}{2g^2_\Lambda} H^2+ \frac{1}{2G_\Lambda} \varphi^2 +\Tilde{V}_\Lambda (H, \varphi) , \\
\Tilde{V}_\Lambda (H, \varphi)
  =& -\frac{1}{4\pi^2} H^2 \left( \ln \frac{H}{\mu^2} -C \right) 
\Big[ \zeta \left( -1, \frac{3}{2}+\frac{X}{2H} \right)
+ \zeta \left( -1, -\frac{1}{2}+\frac{X}{2H} \right) \Big] \notag \\
 &+\frac{1}{4\pi^2} H^2 \Bigg[ \zeta^{(1,0)} \left( -1, \frac{3}{2}+\frac{X}{2H} \right) 
+\zeta^{(1,0)} \left( -1, -\frac{1}{2}+\frac{X}{2H} \right) 
 -2\zeta \left(-1, \frac{1}{2}+\frac{X}{2H} \right) \Bigg] , 
\notag \\ 
X :=& \varphi+\Lambda^2.
\end{align}
Here we have rescaled $H$ as $H \to \frac{1}{g} H$ for later convenience so that the quantum parts $\Tilde{V}_\Lambda$ does not include the $g$ dependence.
  We find that $\tilde{V}_\Lambda(H,\varphi)$ is obtained form $\tilde{V}_\Lambda(H)= \tilde{V}_\Lambda(H,\varphi=0)$ by shifting the variable $\Lambda^2 \to  \Lambda^2+\varphi$:
\begin{equation}
\tilde{V}_\Lambda(H,\varphi) = \tilde{V}_\Lambda(H,\varphi=0)|_{\Lambda^2 \to X }   =   \tilde{V}_\Lambda(H )|_{\Lambda^2 \to X } .
\end{equation}              
The real-valuedness condition for $V_\Lambda$ is replaced by
$
  X-H  >0, \ \text{or}  \ \ H <  X   :=\varphi +\Lambda^2  .
$
In other words, the stability excludes the region:
$
 H \ge     X:=\varphi +\Lambda^2  .
$
Therefore, we define  the \textbf{allowed region for stability}:
\begin{equation}
 \mathcal{R}_\Lambda = \left \{ (H, \varphi) ;  H <  X:=\varphi +\Lambda^2, H \ge0, \varphi> 0 \right \}.
\end{equation}
which is a region below the straight line $ H=  X$ with the slope $1$ and intercept $\Lambda^2$.

$V_\Lambda (H,\varphi)$ can be made real-valued by taking sufficiently large $\Lambda$, as in the case of $V_\Lambda (H)$.
In the absence of $\varphi$, this argument for eliminating the imaginary part does not work in the  small $\Lambda$ region in which  the inequality $ H>\Lambda^2$ is satisfied. This shortcoming is avoided by including $\varphi$. 
In fact,  the allowed region for stability $\mathcal{R}_\Lambda$  becomes narrower for lower value of $\Lambda$, but survives even in the  limit $\Lambda \to 0$.
Hence, the $H$ axis  or $\varphi=0$ is excluded  in the  limit $\Lambda \to 0$.

The running coupling $g_\Lambda$ is monotonically increasing in decreasing $\Lambda$. Therefore,  the tree term $\frac{1}{2} g_\Lambda^{-2} H^2$ also becomes negligible for small enough $\Lambda$.

We can write down the flow equation for $G_\Lambda$.
Solving it, we find that $G^{-1}_\Lambda$   monotonically decreases  as $\Lambda$ decreases. Therefore, the effect of the tree term $\frac{1}{2} G_\Lambda^{-1} \varphi^2$ becomes more and more negligible for smaller $\Lambda$. 
In fact, the increasing of $G_\Lambda$ in decreasing $\Lambda$ is reasonable, since the bound state propagator $G_\Lambda (s)$ will approach the structure with a pole-like dependence on $s$ for small enough $\Lambda$ \cite{Ellwanger94}.
Therefore, the details of the behavior of $G_\Lambda$ does not change the following result qualitatively. 

Thus the existence and location of the minimum can be dominantly determined by the quantum part $\Tilde{V} (H, \varphi)$. 
In view of these, we have looked  for the minimum of $\Tilde{V} _\Lambda (H, \varphi)$ in the region $\mathcal{R}_\Lambda$.

\section{Conclusion and discussion}

We have shown that the Nielsen-Olesen instability of the Savvidy vacuum with homogeneous chromomagnetic condensation is avoided in the framework of the FRG.
Actually, we have shown that the imaginary part of the effective average action vanishes  at sufficiently large infrared cutoff $\Lambda$, and this property can survive at $\Lambda=0$.
This behavior can be understood  as a fixed point solution of the flow equation for the complex-valued effective average action. 
Therefore, the  Nielsen-Olesen instability is an artifact of the  loop calculation in the perturbation theory.

Moreover, we have discussed the physical mechanism for keeping the stability for smaller $\Lambda$: the  stability is maintained even for small $\Lambda$ once  the mass generation occurs  for the off-diagonal gluons (and off-diagonal ghosts). See \cite{Kondo04} for the related works.

The comparison of our result for the effective potential with that of \cite{EGP11}  suggests that  (i)  $H \not= 0$ and $\varphi \not=0$ is realized in the Yang-Mills vacuum. 
Using these solutions \cite{Gies02,EGP11}, moreover, we are able to discuss the possible relationship between the stability and  the scaling/decoupling  solutions which are recently claimed to be  the true infrared solutions in the deep infrared region realizing quark and gluon confinement.
These issues will be further discussed in  future works.

\end{document}